\documentclass{elsart}
\usepackage{graphicx}
\usepackage[fleqn]{amsmath}
\usepackage{amssymb}
\newcommand{\be}{\begin{equation}}
\newcommand{\ee}{\end{equation}}
\newcommand{\bea}{\begin{eqnarray}}
\newcommand{\eea}{\end{eqnarray}}

\allowdisplaybreaks[2]

\begin{document}

\begin{frontmatter}

    
\title{More on the finite size mass shift formula
for stable particles}

\author{Yoshiaki Koma and Miho Koma}

\address{DESY, Theory Group, 
Notkestrasse 85, D-22603 Hamburg, Germany}

\begin{abstract}
The next to leading order (NLO) contribution of 
the generalized finite size mass shift formula
for an interacting two stable particle system 
in a periodic $L^{3}$ box 
is discriminated 
with maintaining its model independent structure 
and validity to all orders in perturbation theory.
The influence of the NLO contribution is examined
for the nucleon mass shift in the realistic nucleon-pion 
system.
\end{abstract}


\end{frontmatter}

\par
Measurements of hadron spectrum in unquenched lattice QCD
simulations always suffer the finite volume effect
from the  associated virtual cloud of lightest 
particles in the spectra, which may
lap the whole lattice once or several times
owing to periodic boundary conditions.
Finite size mass shift formulae, involving the
quantum loop effect of pions in finite volume, 
are thus derived in chiral perturbation theory 
(ChPT)~\cite{Orth:2003nb,Colangelo:2003hf,%
AliKhan:2003cu,Beane:2004tw,Beane:2004ks,%
Detmold:2004ap,Orth:2005kq,Colangelo:2005gd}
and applied to the simulation results
for the purpose of controlling its volume
dependence and of identifying  
the value corresponding to the thermodynamic
limit (see~\cite{Colangelo:2004sc} for a recent review).

\par
In our previous paper, we looked at this 
issue~\cite{Koma:2004wz}
from a  general field theoretical point of view 
(without sticking to ChPT) and derived 
the finite size mass shift formula 
for the interacting two stable particle system
in a periodic $L^{3}$ box,
as an extension of L\"uscher's formula for 
self-interacting bosons~\cite{Luscher:1983rk,Luscher:1986dn}.
Remarkable points of L\"uscher's formula are that 
the finite size mass shift in a periodic box
is related to forward elastic 
scattering amplitudes in infinite volume, which
is model independent,
and can be valid to all orders 
in perturbation theory up to a certain error 
term~\cite{Luscher:1986dn}.

\par
In perturbation theory the physical mass is 
defined from the pole position of the full propagator. 
Using this fact, the finite size mass shift of a (bosonic) 
particle~\footnote{The fermionic mass shift 
can also be defined in a similar way by 
sandwiching the self energies between spinors.}
$\Delta m (L) = M (L) -m$ in Euclidean space
can be defined  as
\bea
\Delta m (L) =
- \frac{1}{2m}[\Sigma_{L}(p)-\Sigma(p)]+O ( (\Delta m)^{2})
\quad \mbox{at}\quad p=(im,\vec{0} \, ) \; ,
\label{eqn:massshift-2boson}
\eea
where $\Sigma_{L}(p)$ and $\Sigma(p)$ denote the self 
energies in the finite and infinite volumes, respectively.
We renormalize the self energy in infinite volume 
as $\Sigma(p)=\frac{\partial}{\partial p^{2}}\Sigma(p)=0$ 
at $p^{2}=-m^{2}$.
In perturbation theory
$\Sigma_{L}(p)$ contains the number of sums over 
discrete spatial loop momenta, 
$\vec{q}_{i}(L) = 2\pi \vec{n}_{i}/L$ $(\vec{n}_{i} \in \mathbb{Z}^{3})$,
depending on the number of loops ($i=1,\ldots, N_{\rm loop}$).
These summations can be rewritten as integrals
by using the Poisson summation formula.
Then the integrand of $\Sigma_{L}(p)$ 
is reduced to the same form as that of $\Sigma(p)$ apart from
the exponential factors $e^{-iL \vec{m}_{i}\cdot \vec{q}_{i}}$
and summations over integer vectors 
$\vec{m}_{i} \in \mathbb{Z}^{3}$.

\par
Since the magnitude of $\vec{m}_{i}\in \mathbb{Z}^{3}$ 
acts as the weight of 
the exponential suppression factor of the 
mass shift formula, the leading order (LO) contribution to 
$\Delta m(L)$
for an asymptotically large $L$ can be specified by 
requiring that only one of them has a non-zero value $|\vec{m}|\equiv 
|\vec{m}_{i}| =1$ 
and the others have zero $|\vec{m}_{j}| = 0$ ($j\ne i$).
In other words, the asymptotic formula can be described by 
the collection of the effective one-loop diagrams
with an exponential factor $e^{-iL \vec{m}\cdot \vec{q}}$, 
where the other loop integrals without exponential factors 
are reduced to the part of the definition of the 
vertex function in infinite volume.
L\"uscher originally discussed this 
case~\cite{Luscher:1983rk,Luscher:1986dn}
and we also did it in the previous paper~\cite{Koma:2004wz}.
The order of the error term in the formula was then
consistent with that of the next to leading order (NLO)
contribution; $|\vec{m}| =  |\vec{m}_{i}| =\sqrt{2}$
and $|\vec{m}_{j}| =0$ ($j\ne i$).

\begin{figure}[t]
\centering\includegraphics[width=12cm]{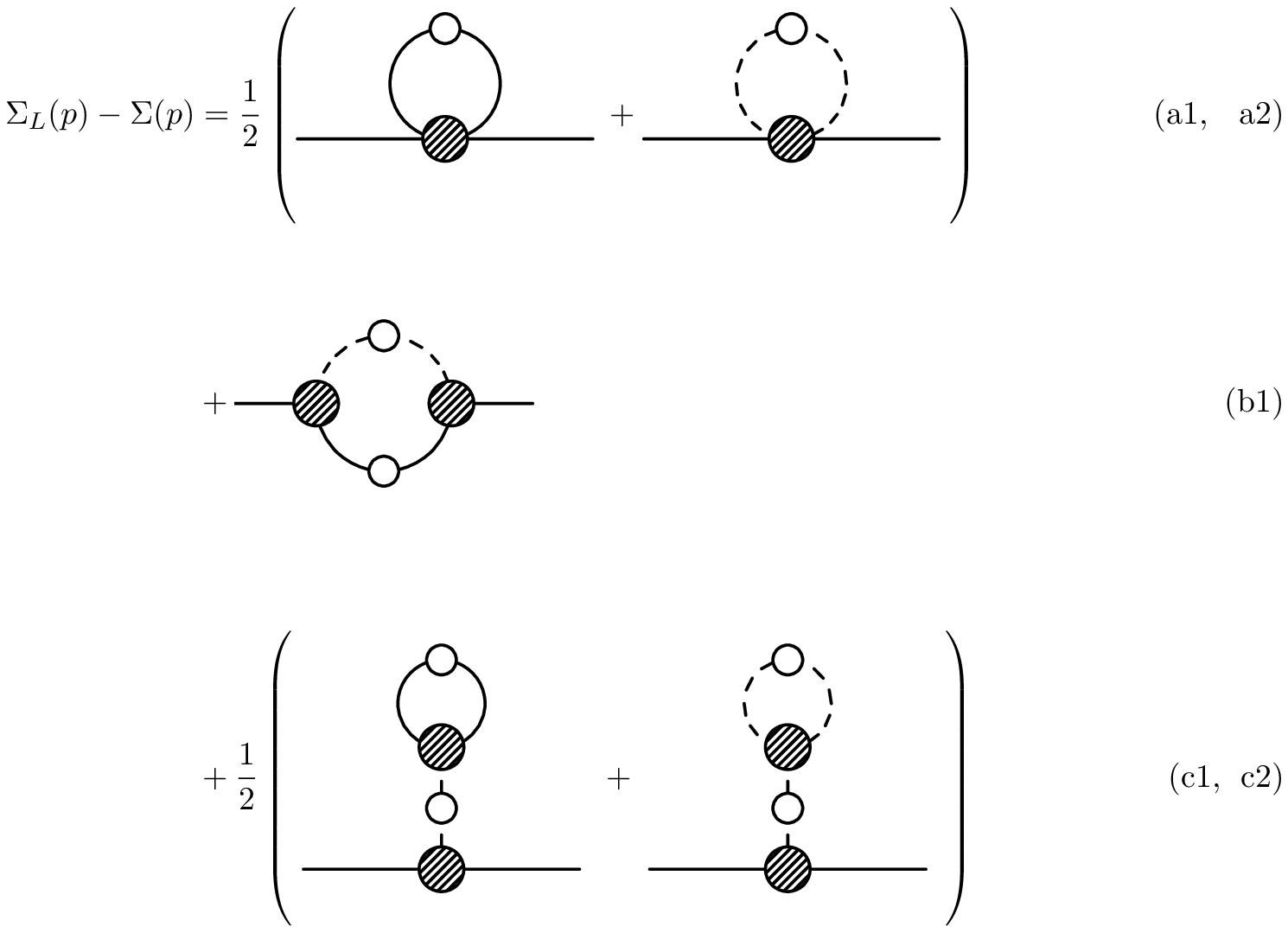}
\caption{Effective one-loop 
self-energy diagrams which contribute to the mass 
shift formula in the  bosonic $A$-$B$ system. 
Solid lines with an empty circle 
correspond to the propagator of $A$ particle, $G_{A}$,
and dashed lines to that of $B$ particle, $G_{B}$. 
Shaded blobs are vertex functions, $\Gamma_{AAB}$, 
$\Gamma_{AAAA}$, $\Gamma_{AABB}$, $\Gamma_{BBB}$, 
at certain orders in perturbation theory.
It is assumed that $A$ particle carries a conserved charge.} 
\label{fig:selfenergy}
\end{figure}

\par
For the realistic application of the formula
to analyzing lattice data, however, it is desirable to 
reduce the ambiguity associated with the error term.
In the present paper, we thus aim to discriminate
the NLO contribution ($|\vec{m}| =\sqrt{2}$)
in the formula for the two particle ($A$-$B$) system,
in particular, while maintaining its model independent structure and
validity to all orders in perturbation theory.
In this case, the task is still the same as for the 
$|\vec{m}| =1$ case; we evaluate the 
effective one-loop diagrams as listed in Fig.~\ref{fig:selfenergy}.
We may here assume that $A$ particle carries a conserved charge,
so that interaction induced  by the three-point vertex $AAB$
and four-point charge conserving vertices are taken into account.
It should be noted that what is nontrivial for such an extension
is not so much evaluating the $|\vec{m}| =\sqrt{2}$ contribution itself
as evaluating the $|\vec{m}| =1$
contribution with an error term at most of the
order of the NNLO contribution ($|\vec{m}| = \sqrt{3}$).
Otherwise the NLO contribution will be obscured 
in the error term.
In fact, it is straightforward to compute the 
$|\vec{m}| =\sqrt{2}$ contribution once the procedure 
is established for $|\vec{m}| =1$.

\par
The result turns out that 
for the mass ratio 
\bea
\alpha \equiv \frac{m_{B}}{m_{A}}
\in (0, \alpha_{\rm max}]
\eea
with $\alpha_{\rm max} \approx 0.418$,
it is possible to discriminate the 
NLO contribution and the final expression 
can be written as
\bea
\Delta m_{A}(L)
&=&
- \frac{1}{16\pi m_{A}}
\left( \frac{6}{L}\right )
\Biggl [
\frac{\lambda^{2}}{2\nu_{B}}e^{-L\mu}
+\int_{-\infty}^{\infty}
\frac{dq_{0}}{2\pi}
e^{-L\sqrt{q_{0}^{2}+m_{B}^{2}}}F_{AB}(iq_{0}) 
\Biggr ]
\nonumber\\*
&&
- \frac{1}{16\pi m_{A}} 
\left( \frac{12}{\sqrt{2}L}\right )
\Biggl [
\frac{\lambda^{2}}{2\nu_{B}}e^{-\sqrt{2}L\mu}
+\int_{-\infty}^{\infty}
\frac{dq_{0}}{2\pi}
e^{-\sqrt{2}L\sqrt{q_{0}^{2}+m_{B}^{2}}}F_{AB}(iq_{0}) 
\Biggr ]
\nonumber\\*
&&
+O(e^{-\sqrt{3}L\mu}) \; ,
\label{eqn:massshift}
\eea
where the first and second lines
correspond to the LO and NLO contributions, respectively.
In this expression, 
\bea
\mu \equiv m_{B}\sqrt{1- \frac{\alpha^{2}}{4}}
= \sqrt{m_{B}^{2}-\nu_{B}^{2}},\qquad
\nu_{B}=\frac{m_{B}^{2}}{2m_{A}} \; ,
\eea
and
$F_{AB}(\nu)$ denotes the forward scattering amplitude 
of $A(p) +B(q) \to A(p) +B(q)$ in infinite volume ($\nu=iq_{0}$ 
the crossing variable).
$F_{AB}(\nu)$ has poles at $\nu=\pm\nu_{B}$.
The coupling $\lambda$ is then defined by exploiting
the residue of $F_{AB}(\nu)$ as
\bea
\lim_{\nu \to \pm \nu_{B}}(\nu^{2}-\nu_{B}^{2})F_{AB}(\nu)=
\frac{\lambda^{2}}{2}\; .
\label{eqn:coupling}
\eea

\par
The basic line of the derivation of Eq.~\eqref{eqn:massshift}
is the same as in our previous paper~\cite{Koma:2004wz},
where the detailed notation of 
the propagators and the vertex functions are also given.
In what follows, we shall present a derivation for the case that both 
$A$ and $B$ particles are bosons.
The extension to the case that particle $A$ is a fermion 
is straightforward, and the final result is exactly the same as in 
Eq.~\eqref{eqn:massshift}, which is one of the advantages 
of the model independence of the formula.
Here, we concentrate on evaluating the diagram
(b1) in Fig.~\ref{fig:selfenergy}, which is typical for the
two particle system and provides us with a key idea how to 
control the error term.
The other diagrams are then evaluated in a similar way.

\par
The self-energy diagram (b1) for $|\vec{m}|=1$ is expressed as
\bea
I_{b1}^{(|\vec{m}|=1)}  &=& 
6
\int  \frac{d^{4}q}{(2 \pi)^{4}} 
e^{iLq_{1}}
\Gamma_{\mathit{AAB}}(-p, (1-\eta)p+q ;  \eta p-q) 
G_{A}( (1-\eta)p+q) 
 \nonumber\\*
&& \times 
G_{B}( \eta p - q) 
\Gamma_{\mathit{AAB}}
(p, -(1-\eta)p-q ; -\eta p + q)  \biggr |_{p=(im_{A},
\; \vec{0} \, )}\; ,
\label{eqn:integral-b1}
\eea 
where $\eta$ is a real parameter
at least in the range $[0,2 \delta (\alpha)]$
for $\delta (\alpha)\in (0,1/2]$.
For the purposes of 
evaluating the integral
$\eta$ can be chosen appropriately depending on the 
mass ratio $\alpha$.
Our concern is whether there exists such a set
of $\eta$ and $\delta$ for a given $\alpha$ to make
the error terms smaller than the desired order of magnitude.
For our purpose this is $O(e^{- L \gamma\mu})$ 
with $\gamma=\sqrt{3}$.
In our previous work~\cite{Koma:2004wz}, we chose
$\eta=\delta=\alpha^{2}/2$, which was sufficient to
control the error term up to $O(e^{- \sqrt{2} L \mu})$.
But this choice turns out to be inappropriate 
in the present case (see our final choice in 
Eq.~\eqref{eqn:optimal-eta-delta}).
The overall factor 6 originates from rotational invariance among
$q_{1}$, $q_{2}$ and $q_{3}$.

\par
Firstly, we perform the complex $q_{1}$ 
contour integration by focusing on 
the poles of the propagators $G_{A}$ and 
$G_{B}$ of one-particle states at~\footnote{We 
are assuming that there is no
bound state below the two-particle threshold.} 
\bea
&&
q_{1}^{(A)}=i\sqrt{ q_{0}^{2}+q_{\perp}^{2}+ 
(2 \eta -\eta^{2}) m_{A}^{2}
+ i 2 (1-\eta) m_{A}q_{0}} \; , \\
&&
q_{1}^{(B)}=i\sqrt{q_{0}^{2}+q_{\perp}^{2}+
m_{B}^{2}-\eta^{2}m_{A}^{2}
- i 2\eta m_{A} q_{0}} \; ,
\eea
in the complex $q_{1}$ upper half plane, 
respectively,~\footnote{For $a$, $b \in \mathbb{R}$, 
\bea
\mbox{Im}~(i\sqrt{a+ib}) = 
 \sqrt{(\sqrt{a^{2}+b^{2}}+a )/2} \;.
\eea
}
where $q_{\perp}^{2} = q_{2}^{2}+q_{3}^{2}$.
We may set a contour which goes along the real $q_{1}$ 
line and the line $\mbox{Im}~q_{1}=\theta_{1}>0$
closed at $\pm \infty$ to 
pick up the residues at $q_{1}^{(A)}$ and/or $q_{1}^{(B)}$.
In order to relate the mass shift to 
the on-shell forward scattering amplitude
like in Eq.~\eqref{eqn:massshift}, we find at this point that 
the upper path $\theta_{1}$ must be chosen 
so as to satisfy the following conditions;
\begin{enumerate}
\renewcommand{\labelenumi}{(\roman{enumi})}
\item the contribution from the upper path itself is 
smaller than the error term $O(e^{-  L \gamma\mu})$,
\item  the contour  covers the range of $\mbox{Im}~q_{1}^{(A)}$ 
and/or $\mbox{Im}~q_{1}^{(B)}$ 
for $q_{0}$ and $q_{\perp}$ in a certain ball
\bea
\mathbb{B} = 
\{ (q_{0},q_{\perp})\in \mathbb{R}^{3}~|~q_{0}^{2}+q_{\perp}^{2}
\leq 
\nu^{2}
\} \; ,
\label{eqn:ball}
\eea
\item  the  contour picks up no residue except for
the poles at $q_{1}^{(A)}$ and/or $q_{1}^{(B)}$.
\end{enumerate}
Here the condition (iii) 
must be guaranteed even if $q_{0}$ is extended to a complex variable
and shifted as $q_{0}\to q_{0}-i(1-\eta)m_{A}$
for $q_{1}^{(A)}$ and/or $q_{0}\to q_{0}+i \eta m_{A}$
for $q_{1}^{(B)}$, where $q$ satisfies the on-shell condition
$q^{2} = -m_{A}^{2}$ and/or $q^{2}=-m_{B}^{2}$.

\par
\begin{figure}[t]
\centering\includegraphics[width=13cm]{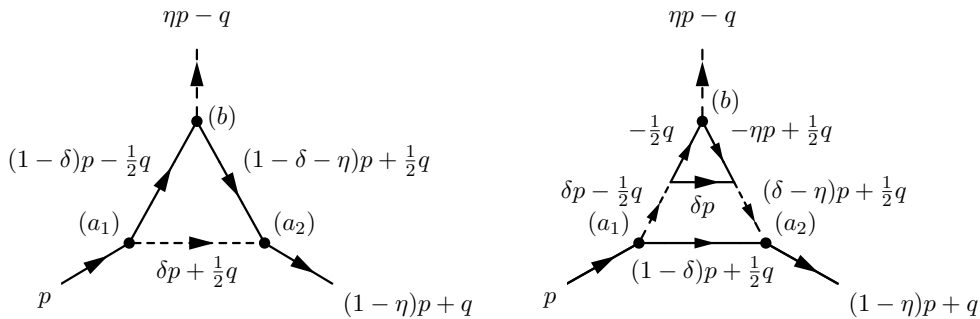}
\caption{The external momentum flow 
in the $AAB$ vertex function. 
Arrows represent the flow direction.}
\label{fig:aab-vertex}
\end{figure}

To examine the condition (iii),
we use the fact that the vertex function 
$\Gamma_{\mathit{AAB}} 
( \mp p, (1-\eta)p \pm q ;  \eta p \mp q)$ 
with $\eta \in [0,2\delta]$,
initially defined for $(p,q) \in \mathbb{R}^{4}
\times \mathbb{R}^{4}$,
is analytic inside the complex domain
\bea
\mathbb{D} &=& 
\{(p,q) \in \mathbb{C}^{4}\times \mathbb{C}^{4} ~|~ 
\left (
\mathrm{Im} \{ (1-\delta)\, p \pm \frac{1}{2} \,q \} 
\right )^{2} < m_{A}^{2}\;,\nonumber\\*
&&\qquad\qquad\qquad\qquad\quad
\left(\mathrm{Im} \{ \delta\, p \pm \frac{1}{2} \,q \} 
\right)^{2} < m_{B}^{2}\; \} \; .
\label{eqn:analyticity}
\eea
The basic observation for finding this domain is that 
the vertex function at any higher order in perturbation 
theory consists of a set of $A$ and $B$ lines
(free propagators).
The denominator of the $l$th $A$ or $B$ line is then parametrized as
$(k(l)+r(l))^{2}+m_{A}^{2}$ or  $(k(l)+r(l))^{2}+m_{B}^{2}$, 
where $k(l)$ is the  external momentum flow given by
a combination of complex variables $p$ and $q$, and $r(l)$ is 
a combination of internal loop momenta to be integrated out, 
which is a real variable in Euclidean space.
It then follows that the vertex function has no singularity if 
$(\mathrm{Im}~k(l))^{2} < m_{A}^{2}$ 
and $(\mathrm{Im}~k(l))^{2} < m_{B}^{2}$ for all $A$ and $B$ lines.
In order to find the possible choices of $k(l)$,
we may label the three bare vertices where
the external momenta, $p$, $(1-\eta)p + q$ and $\eta p  - q$,
are plugged in (and out) as $a_{1}$, $a_{2}$ and $b$, respectively
(e.g. $a_{1}=a_{2}=b$ at the tree level).
Note that whenever $A$ particle carries a conserved charge, 
there always exists a set of $A$ lines connecting $a_{1}$ and $a_{2}$.
In Fig.~\ref{fig:aab-vertex}, we show the possible
external momentum flow inside the $AAB$ vertex function;
they are basically classified into two cases,
the connected $A$ lines flow through $b$ (left) and
the connected $A$ lines do not flow through $b$ (right).
One can add any internal lines depending on the order
in perturbation theory, which however 
carry no external momentum and
do not affect the singularity of the vertex function.
Inserting the largest external momenta for $A$
and $B$ lines into $(\mathrm{Im}~k(l))^{2} < m_{A}^{2}$ 
and $(\mathrm{Im}~k(l))^{2} < m_{B}^{2}$, respectively,
one can specify the domain $\mathbb{D}$ 
as in~\eqref{eqn:analyticity}.

\par
We then realize that there is no integration path $\theta_{1}$ 
at any value of $\alpha$  
which satisfies all conditions (i)$\sim$(iii) for both
poles $q_{1}^{(A)}$ and $q_{1}^{(B)}$ simultaneously.
However, we find that it is possible to choose
$\theta_{1}=\sqrt{\gamma^{2}\mu^{2}+\eta^{2}m_{A}^{2}}$
with the upper bound of the ball
$\nu^{2}=\gamma^{2} 
\mu^{2} - m_{B}^{2}+\eta^{2}m_{A}^{2}$
in Eq~\eqref{eqn:ball}, 
which satisfies the conditions
only for $q_{1}^{(B)}$ within a limited range of~$\alpha$.
The choice of $\eta$ and $\delta$ is quite subtle, but
by choosing 
\bea
\eta = c \; \delta \; ,\qquad
\delta = 
\alpha  \sqrt{ 
\frac{4- \gamma^{2} (1-\frac{\alpha^{2}}{4})}{
2(c^{2}+2c+2)}} 
\; . 
\label{eqn:optimal-eta-delta}
\eea
with $c=0.95$, the allowed range of $\alpha$ is 
maximized as $\alpha \in (0, \alpha_{\rm max}]$, where
$\alpha_{\rm max}\approx 0.418$.~\footnote{In this region 
the contribution from $q_{1}^{(A)}$
can be neglected since it is of $O(e^{- L \gamma\mu})$.
To show this explicitly, we need to carry out the
$q_{0}$ contour integration by performing 
the momentum shift $q_{0} \to q_{0}+ i \theta_{0} m_{A}$
for $q_{1}^{(A)}$.
Here, the domain $\mathbb{D}$ constrains
$\theta_{0}$ to be
$\theta_{0} < ((4-c)\delta-2(1-c^{2}/4)\delta^{2})/(1-(2-c)\delta)$,
$\theta_{0} > -((4-c)\delta-2(1-c^{2}/4)\delta^{2})/(3-(2+c)\delta)$,
$\theta_{0} > -(2 \alpha^{2} -c \delta 
-2(1-c^{2}/4)\delta^{2})/(1+(2-c)\delta)$, and
$\theta_{0} > -(2 \alpha^{2} -c \delta 
-2(1-c^{2}/4)\delta^{2})/(1-(2+c)\delta)$.
On the other hand, we find $\mbox{Im}~q_{1}^{(A)} \geq \gamma \mu$,
if $\theta_{0} < \eta - 1 + \sqrt{1-\gamma^{2}\alpha^{2}
(1-\alpha^{2}/4)}$.
Solving these inequalities (numerically) with 
Eq.~\eqref{eqn:optimal-eta-delta},
we find that $c=0.95$ yields the maximal value of $\alpha$.}
Thus we obtain
\bea
I_{b1}^{(|\vec{m}|=1)} 
&=&
6 i \int_{\mathbb{B}}\frac{dq_{0}d^{2}q_{\perp}}{(2 \pi)^{3}} 
\frac{e^{i L q_{1}}}{2 q_{1}}
 \Gamma_{\mathit{AAB}}(-p, (1-\eta)p+q;  \eta p -q)  
 \nonumber\\*
&&
\times G_{A}((1-\eta)p+q)
\Gamma_{\mathit{AAB}}(p,-(1-\eta )p-q; - \eta p +q) 
|_{q_{1}=q_{1}^{(B)}} 
\nonumber\\*
&&
+ O(e^{-L\gamma \mu})\; .
\label{eqn:integral-b1-2}
\eea
We remark that if we set $\gamma = 2$, where
$O(e^{-2L\mu})$ corresponds to the NNNLO contribution,
there is no parameter set $(\theta_{1},\eta, \delta,\nu)$ which 
fulfills the above conditions at any value of $\alpha$,
indicating that one cannot discriminate
the NNLO contributions along this line.

\begin{figure}[t]
\centering\includegraphics[width=8cm]{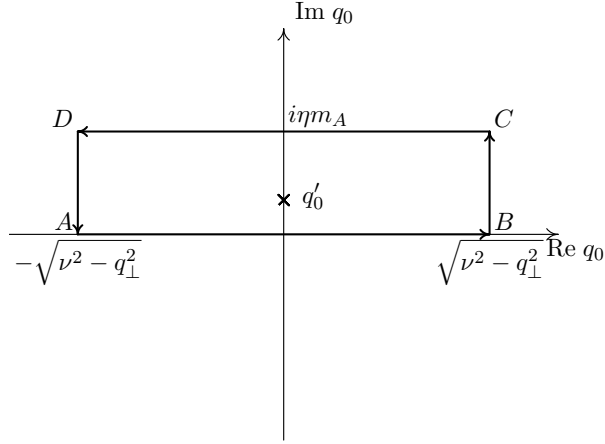}
\caption{$q_{0}$ integration contour.}
\label{fig:q0int2}
\end{figure}

\par
\begin{figure}[t]
\centering\includegraphics[width=12.5cm]{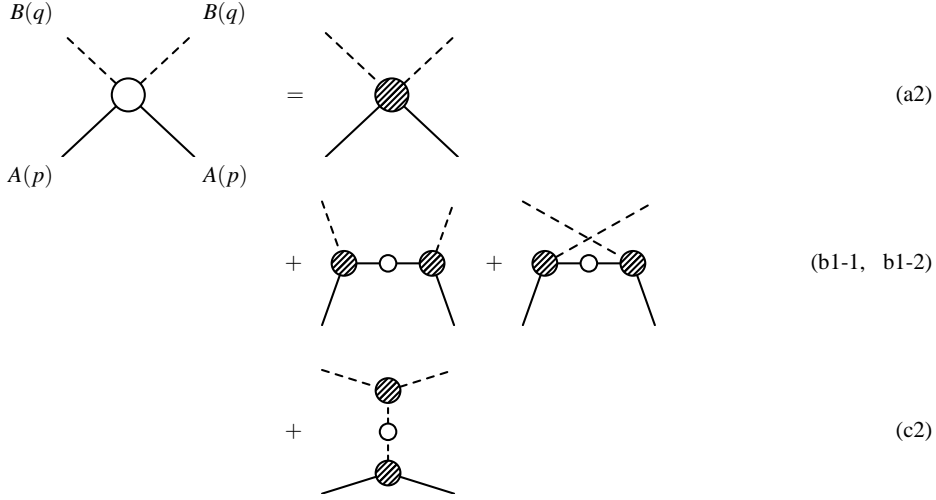}
\caption{Ingredients of $F_{\mathit{AB}}(\nu)$
in the $A$-$B$ system.
The labels represent the correspondence with 
the self-energy diagrams in Fig.~\ref{fig:selfenergy}.}
\label{fig:ab-scattering}
\end{figure}

Secondly, we perform the complex $q_{0}$ contour integration 
along the path in Fig.~\ref{fig:q0int2}.
The path $CD$ is parametrized by shifting the momentum 
$q_{0} \to q_{0}+i\eta m_{A}$.
On this path the argument of $G_{A}$ becomes
$p+q$, where $q$ satisfies the on-shell condition
$q^{2}=-m_{B}^{2}$,
since $q_{1}^{(B)} = i\sqrt{q_{0}^{2}+q_{\perp}^{2}+m_{B}^{2}}$.
Inside the contour the integrand has no pole except at
\bea
q_{0}= q_{0}'=im_{A} \left (\eta - \frac{\alpha^{2}}{2} 
\right ) \; ,
\eea
which is guaranteed by the above condition (iii).
Note that $0 < \mbox{Im}~q_{0}' < \eta m_{A}$ for the value of 
$\eta$ specified by Eq.~\eqref{eqn:optimal-eta-delta}.
The contributions from the paths $BC$ and $DA$ are at most
$O(e^{-L\gamma \mu})$ since 
$\mbox{Im}~q_{1}^{(B)} \geq \gamma \mu$, which is 
due to the choice of the ball $\mathbb{B}$ 
with $\nu^{2}=\gamma^{2} \mu^{2} - m_{B}^{2}+\eta^{2}m_{A}^{2}$
in~\eqref{eqn:ball}.
Thus the integral~\eqref{eqn:integral-b1-2} is
replaced by one along the path $CD$ with 
the residue contribution at $q_{0}'$
\bea
I_{b1}^{(|\vec{m}|=1)} 
&= &
3  
\int_{\mathbb{B}'} \frac{d^{2}q_{\perp}}{(2 \pi)^{2}}
\frac{e^{- L\sqrt{ q_{\perp}^{2} + \mu^{2} }
}}{2\sqrt{ q_{\perp}^{2}+\mu^{2}}} \;
\frac{\lambda^{2}}{2\nu_{B}}
\nonumber\\*
&&
+ \; 
 6  \int_{\mathbb{B}} \frac{dq_{0}d^{2}q_{\perp}}{(2 \pi)^{3}}
\frac{e^{-L\sqrt{q_{0}^{2}+q_{\perp}^{2}+m_{B}^{2}}}}
{2\sqrt{q_{0}^{2}+q_{\perp}^{2}+m_{B}^{2}}}
\;  F_{\mathit{AB}}^{(b1-1)}(iq_{0}) 
+O(e^{-L \gamma \mu}) \; ,
\label{eqn:ib1m}
\eea
where
\bea
F_{\mathit{AB}}^{(b1-1)}(iq_{0}) 
&=&
\Gamma_{\mathit{AAB}}(-p, p+q; -q)  G_{A}(p+q)
 \nonumber\\*
 &&
 \times
\Gamma_{\mathit{AAB}}(p, - p-q ; q) 
|_{p^{2}=-m_{A}^{2},\; q^{2}=-m_{B}^{2}} \; 
\eea
corresponds to a one-particle-irreducible (1PI) 
part of the forward scattering amplitudes
of $F_{\mathit{AB}}(\nu=iq_{0})$, 
graphically represented as (b1-1) in Fig.~\ref{fig:ab-scattering}.
By using the crossing relation
$F_{\mathit{AB}}^{(b1-1)}(-\nu)=F_{\mathit{AB}}^{(b1-2)}(\nu)$,
one can replace
$F_{\mathit{AB}}^{(b1-1)}(\nu)$  in Eq.~\eqref{eqn:ib1m}  by
$(F_{\mathit{AB}}^{(b1-1)}(\nu)+F_{\mathit{AB}}^{(b1-2)}(\nu))/2$.
In the first term in Eq.~\eqref{eqn:ib1m},
the effective renormalized coupling $\lambda$ 
is defined by Eq.~\eqref{eqn:coupling}, where
$F_{\mathit{AB}}^{(b1-1)}(\nu)$ or 
$F_{\mathit{AB}}^{(b1-2)}(\nu)$
has the pole at $\nu=\pm \nu_{B}$,
and the integral region 
is specified by inserting $q_{0} = q_{0}'$
to $\mathbb{B}$:
\bea
\mathbb{B}' = \{ q_{\perp} \in \mathbb{R}^{2}
~|~ q_{\perp}^{2} \leq   (\gamma^{2}-1)\mu^{2}
+2\eta m_{A}^{2}(\eta -\frac{\alpha^{2}}{2})\} \; .
\eea

\par
We then carry out the $q_{\perp}$ integration 
in Eq.~\eqref{eqn:ib1m} by using the integral formula
\bea
\int_{-\infty}^{\infty} \frac{d^{2}q_{\perp}}{(2 \pi)^{2}}
\frac{e^{- L\sqrt{q_{\perp}^{2}+ \rho^{2}}}}
{2 \sqrt{q_{\perp}^{2}+\rho^{2}}}
=
\frac{1}{4\pi L}e^{- L \rho} \; ,
\label{eqn:formula_qperp_int}
\eea
where the integral region
can be extended from $\mathbb{B}$ or $\mathbb{B}'$ to infinity, 
because the boundary contributions of $\mathbb{B}$ and $\mathbb{B}'$
are already smaller than the order of the error term.
Hence, we end up with
\bea
I_{b1}^{(|\vec{m}|=1)} 
&= &
\frac{1}{8\pi}
\left ( \frac{6}{L}\right )
\Biggl [
\frac{\lambda^{2}}{2 \nu_{B}} e^{-L  \mu }
+
\int_{-\infty}^{\infty}\! \!  \frac{dq_{0}}{2 \pi}\;
e^{-L\sqrt{q_{0}^{2}+m_{B}^{2}}} \; 
\{ F_{\mathit{AB}}^{(b1-1)}(iq_{0})+F_{\mathit{AB}}^{(b1-2)}(iq_{0}) \}
\Biggr ]  
\! \!  
 \nonumber\\*
 && 
+ O(e^{- \sqrt{3}L\mu})
\; .
\label{eqn:boson-result}
\eea
Other self-energy diagrams in Fig.~\ref{fig:selfenergy} 
can be evaluated in a similar way up to $O(e^{- \sqrt{3}L\mu})$,
yielding the corresponding 1PI part of the forward scattering amplitude.
Note that the contributions from  the 
self-energy diagrams (a1) and (c1) are already smaller than
$O(e^{- \sqrt{3}L\mu})$ for $\alpha \in (0,\alpha_{\rm max}]$.
By combining all contributions we can
discriminate the $|\vec{m}|=1$ contribution 
up to $O(e^{- \sqrt{3}L\mu})$.

\par
The $|\vec{m}|=\sqrt{2}$ contribution 
is given by the integral
\bea
I_{b1}^{(|\vec{m}|=\sqrt{2})} &=& 
12
\int  \frac{d^{4}q}{(2 \pi)^{4}} 
e^{iL(q_{1}+q_{2})}
\Gamma_{\mathit{AAB}}(-p, (1-\eta)p+q ;  \eta p-q) 
G_{A}( (1-\eta)p+q) 
 \nonumber\\*
&& \times 
G_{B}( \eta p - q) 
\Gamma_{\mathit{AAB}}
(p, -(1-\eta)p-q ; -\eta p + q)  \biggr |_{p=(im_{A},
\; \vec{0} \, )}\; .
\label{eqn:integral-b1-next}
\eea 
Rotating the $q_{1}$-$q_{2}$ axis by $\pi/4$, we define
$\tilde{q}_{1} = (q_{1}+q_{2})/\sqrt{2}$ and
$\tilde{q}_{2} = (-q_{1}+q_{2})/\sqrt{2}$.
Then, apart from the new exponential factor 
$e^{i\sqrt{2}L\tilde{q}_{1}}$ and an overall factor 12, 
the integrand becomes
exactly the same as in Eq.~\eqref{eqn:integral-b1}.
Thus the evaluation is  straightforward and the result is
\bea
I_{b1}^{(|\vec{m}|=\sqrt{2})} 
&= &
\frac{1}{8\pi}
\left ( \frac{12}{\sqrt{2}L}\right )
\Biggl [
\frac{\lambda^{2}}{2 \nu_{B}} e^{-\sqrt{2} L  \mu }
 \nonumber\\*
 && 
+
\int_{-\infty}^{\infty}\! \!  \frac{dq_{0}}{2 \pi}\;
e^{-\sqrt{2}L\sqrt{q_{0}^{2}+m_{B}^{2}}} \; 
\{ F_{\mathit{AB}}^{(b1-1)}(iq_{0})+F_{\mathit{AB}}^{(b1-2)}(iq_{0}) \}
\Biggr ]  
\! \!  
\nonumber\\*
&& 
+O(e^{- \sqrt{6} L\mu})
\; ,
\label{eqn:boson-result-2}
\eea
where the  error term becomes automatically 
smaller than the $|\vec{m}|=1$ case.
Evaluating other diagrams similarly and combining the result of
$|\vec{m}|=1$, we arrive at 
the mass shift formula in Eq.~\eqref{eqn:massshift}.

\par
Finally, let us  examine the influence of the 
$|\vec{m}|=\sqrt{2}$ contribution
by looking at the nucleon mass shift 
in the realistic $N$-$\pi$ system 
with $m_{N}=938$ MeV and $m_{\pi}=140$ MeV.
As the mass ratio is $\alpha=m_{\pi}/m_{N}=0.149$,
Eq.~\eqref{eqn:massshift} is applicable.
Moreover, since the formula is valid to all orders in perturbation 
theory and is expected to hold nonperturbatively, 
we may insert the empirical 
$N$-$\pi$ scattering amplitude into Eq.~\eqref{eqn:massshift}.
The subthreshold expansion of the $N$-$\pi$ forward scattering 
amplitude around $\nu = 0$ is parametrized as~\cite{Hohler:1984ux}
\bea
F_{\mathit{N\pi}}(\nu) = 6 m_{N} D^{+}(\nu)\; ,
\label{eqn:n-pi-famp}
\eea
where
\bea
D^{+} (\nu)
= 
\frac{g^{2}}{m_{N}}\frac{\nu_{B}^{2}}{\nu_{B}^{2}-\nu^{2}}
+d_{00}^{+} \; m_{\pi}^{-1} +d_{10}^{+} \; m_{\pi}^{-3} \nu^{2}
+d_{20}^{+} \; m_{\pi}^{-5} \nu^{4} + O(\nu^{6}) \; .
\label{eqn:d-plus}
\eea
The isospin sum is taken in Eq.~\eqref{eqn:n-pi-famp}, 
neglecting the effect of isospin symmetry breaking.
The coupling constant is $g^{2}/4 \pi =14.3$.
The first term in Eq.~\eqref{eqn:d-plus}
is identified with the pseudovector nucleon Born term
with $\nu_{B}=m_{\pi}^{2}/2m_{N} \approx 0.07 m_{\pi}$.
The effective coupling is then easily computed 
by using Eq.~\eqref{eqn:coupling} as
$\lambda^{2}=-12g^{2}\nu_{B}^{2}$.
The coefficients of the other terms are given by 
$d_{00}^{+} = -1.46(10)$, $d_{10}^{+}=1.12(2)$ and 
$d_{20}^{+} = 0.200(5)$~\cite{Hohler:1984ux}.
Hereafter we only take into account the mean of these values.

\begin{figure}[t]
\centering\includegraphics[width=13cm]{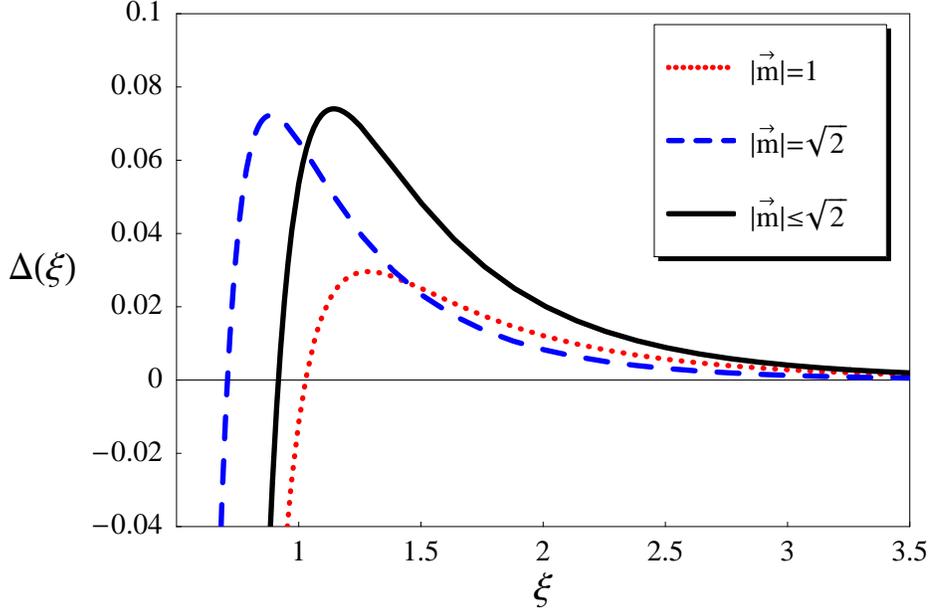}
\caption{The nucleon mass shift as a function of $\xi=Lm_{\pi}$.} 
\label{fig:delta_xi}
\end{figure}

\par
In Fig.~\ref{fig:delta_xi}, we plot
$\Delta (\xi =Lm_{\pi}) \equiv \Delta m_{N}/m_{N}$ 
for the $|\vec{m}|=1$ and $|\vec{m}|=\sqrt{2}$ 
contributions (dotted and dashed lines, respectively) and
the sum of these contributions as $|\vec{m}| \leq \sqrt{2}$
(solid line), where $\xi =1$ corresponds to $L=1.4$~fm.
It reveals that the $|\vec{m}|=\sqrt{2}$ contribution is 
quite large for the plotted region of $\xi$.
For instance at $\xi = 2$ the mass shift is 
expected to occur more than 1.2 ($|\vec{m}|=1$) 
+ 0.8 ($|\vec{m}|=\sqrt{2}$) = 2.0~\% ($\gtrsim$ 20~MeV).
If one estimates the $|\vec{m}| =\sqrt{3}$ contribution
itself at $\xi = 2$,~\footnote{This is easily computed
as the $|\vec{m}| = \sqrt{2}$ case, although the order 
is consistent with the error term.}
this merely
contributes the mass shift  by 0.2~\%.
This is due to the smaller geometrical factor 
(e.g. $6$ for $|\vec{m}|=1$, $12/\sqrt{2}\approx 8.49$ 
for $|\vec{m}|=\sqrt{2}$,
and $8/\sqrt{3}\approx 4.62$ for $|\vec{m}|=\sqrt{3}$)
as well as the larger exponential decay factor.
In this sense the nucleon mass shift formula 
is significantly modified by 
discriminating the NLO contribution.
Note that the negative mass shift for $\xi \lesssim 1$ is 
due to the contribution from the term involving 
the $N$-$\pi$ forward scattering 
amplitude (ingredients of the  $|\vec{m}| =1$ curve
can be found in Ref.~\cite{Koma:2004wz}).

\par
To summarize, we have investigated the 
finite size mass shift formula 
for the two stable particle system
in a periodic $L^{3}$ box.
We have found that it is possible 
for the mass ratio $\alpha \in (0,\alpha_{\rm max}]$
with $\alpha_{\rm max} \approx 0.418$
to discriminate the NLO contribution
with maintaining its model independent structure and
validity to all orders in perturbation theory.
The final expression is then written as in Eq.~\eqref{eqn:massshift}.
Along the way we have also realized that
it is impossible to discriminate the 
NNLO contribution along the line discussed above
once the error term is set by~$\gamma=2$.
In fact, in order to discriminate more higher
order contributions, one should go back to the 
definition of the mass shift in 
Eq.~\eqref{eqn:massshift-2boson}.

\par
We are grateful to the members of lattice forum 
in DESY theory group in Hamburg, 
in particular, H.~Wittig and 
I.~Montvay for valuable discussions and comments.
We appreciate fruitful comments from P.~Weisz.
Y.K. thanks T.R.~Hemmert for useful discussions 
at the meeting of the DFG Forschergruppe `Lattice Hadron
Phenomenology' at DESY-Zeuthen in February.

\end{document}